# Using Cyber Threat Intelligence to Prevent Malicious Known Traffic in a SDN Physical Testbed

Jorge Buzzio García, Víctor Salazar Vilchez, Jeffrey Zavala Castro, Jose L. Quiroz Arroyo
Grupo de Redes Avanzadas y Ciberseguridad (G-RAC)
Instituto Nacional de Investigación y Capacitación de Telecomunicaciones, Universidad Nacional de Ingeniería INICTEL-UNI
Lima - Perú

*Abstract*—Since the use of applications and communication tools has increased, one of the concerns of the responsible for network security has been to protect information and information systems, as well as to provide trust to end users for the use of information and communication technologies. Nowadays, attacks on the network have increased and undergone modifications, which make the task for traditional security devices difficult, being necessary to add the intelligence to face the new attacks generated in the network. Hence the need to incorporate Cyber Threat Intelligence (CTI) as a new component in the network. This work focuses on the use of information provided by a CTI to improve the security of Software Defined Networks (SDN), and at the same time, analyze how malicious traffic could be blocked in a physical testbed.

*Keywords—CTI, CIF, SDN, Controller, OpenFlow, Zodiac.*

## I. INTRODUCTION

A sustained annual growth of cyberattacks perpetrated worldwide, the greater complexity for the detection and subsequent mitigation of cyber-incidents, added to the shorter times for them to get a global reach; they are some of the main characteristics of the current context of cyberspace; all the aforementioned is reflected in the world economy, where the estimated amount lost, generated by cybercrimes for 2018, reached 600 billion dollars [1].

Having a panorama so complicated for Computer Emergency Response Teams (CERT / CSIRT), security officers and system administrators, it is essential to provide the greatest amount of resources and information necessary, so that they can identify and prevent threats to which your organization may be exposed. In this sense, CTI has emerged as added value that has been increasingly accepted and used by organizations for the detection of threats and attacks, response to incidents, vulnerability management, blocking of threats; organizations make use of relevant information by playing the role of consumer and/or data provider CTI [2].

Even more, it is added the inconveniences that characterize the management of a network based on the traditional model, where security policies must be implemented in the network devices (router, switches), even in middle boxes (firewall, IDS, IPS), which are configured through particular commands that differ between manufacturers. In such a way, this article presents the incorporation of CTI taking as test scenario an implementation on real hardware of a SDN, an emerging network architecture whose adoption in recent years has been growing considerably [3], which is characterized by the separation of the control and data planes located in the network devices, unifying and moving the logic of control to an entity that is called the SDN Controller, thus enabling the programmability of the network through the development of software-based applications, simplifying management and decreasing the probability of human error in the implementation or modification of security policies [4].

The following sections of this article are organized as follows: In Section II, we review theoretical information and the most relevant related works. In Section III, we describe our proposal. In section IV, we present and discuss the results. Finally, in Section V we show the conclusions.

## II. BACKGROUND & RELATED WORKS

In [5], the authors used CTI and an intrusion detection system (BroIDS), in order to provide a proactive solution; the counter measures for attacks and threats are converted into flow entries that are entered into the affected switches. Furthermore, if a new attack is detect, in addition to blocking it, it is redirected to a Honeypot for further analysis. This study does its test on a virtual scenario. In [6], the authors proposed a model that allows defending against cyber threats in SDN based on risk analysis, in order to be implemented in production environments. The model collects data from CTI TAXII servers, which is analyzed and those with lower reliability are filtered, then it is determined what actions to take in the SDN network in order to prevent attacks. This model is proved also on a virtual scenario. However, these two works do not contemplate a physical testbed and an efficient use of the flow tables of the SDN switches (pipeline process) for the flow entries that will be installed, since only use one flow table, so the packet will take more time to perform an action.

### A. Software Defined Network

The Software Defined Networks (SDN) allow a centralized control of the network, since the data plane and the control plane are decoupled. The data plane is responsible for forwarding packets, while the control plane provides intelligence in the design of routes and Quality of Service (QoS), among other things, in order to cope with changing traffic patterns. The communication between the SDN controller (control plane) and the SDN switches (data plane) is given by the OpenFlow protocol using a secure channel. [7]

### B. OpenFlow

The OpenFlow protocol allows the controller to manage the flow entries, these inputs have as main components the priority fields, matching fields, counters and a set of actions and instructions to be carried out with the matched packets. There are three types of OpenFlow messages: Controller-to-Switch messages that allow the controller to manage the logical state of the switch, such as configuration and flow entries. Asynchronous messages used to inform the controller about changes in the network or the state of the switch. Symmetric messages are used when the connection is established for the first time, to measure the bandwidth, the status of a connection or to verify if the device is operational. In this work, OpenFlow protocol version 1.3.0 is used due to the support of the pipeline process, in which the process of matching search with an incoming packet is done by taking into account multiple OpenFlow tables. [8]

*C. Zodiac FX*

The Zodiac Fx switch was designed to be used for researchers and students who want to implement a software defined network in a laboratory environment. Developed by the company Northbound Networks, it supports versions 1.0 and 1.3 of the OpenFlow protocol, it has four Ethernet ports, an USB port through which enters the command line interface, and also provides power for its operation and has an Amtel SAM4E8CA processor. [9]

*D. Cyber Threat Intelligence*

CTI collects information from reliable sources providing the ability to recognize suspicious and malicious activities, thus it helps making decisions to prevent damage to the organization. Initial surveys [10], showed the tendency of use and investment of resources for the implementation of solutions based on CTI.

Nowadays, an increase in the acceptance and application of CTI in organizations is confirmed with documented requirements for its use and implementation of good practices [11]. However, still there are a wide variety of industries that claim not have them yet.

The threat intelligence used is CIF (Collective Intelligence Framework), which obtains "feeds" type information from external and internal sources that can be shared among organizations. This information is characterized by being filtered from the observable type, which could be: IPv4, FQDN, URL, IPv6, email.

III. PROPOSE DESIGN

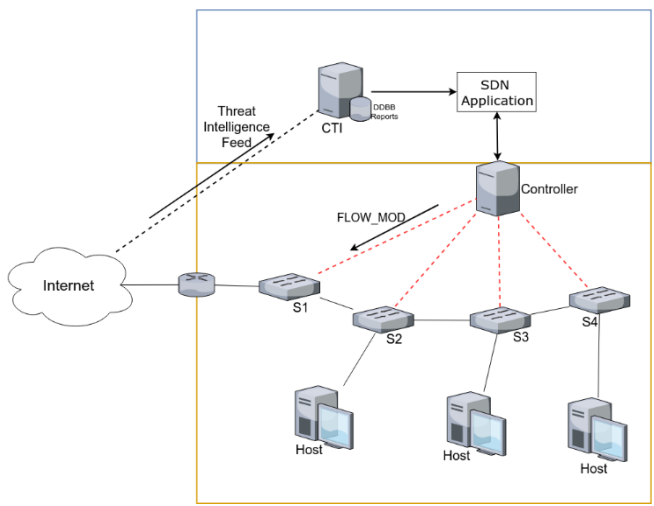

Fig. 1. Testbed Topology

The proposed testbed is composed by five elements: the CIF server, SDN controller (OpendayLight), SDN application, OpenFlow switch (Zodiac FX) and the hosts. The CIF server and the SDN Controller each use a dedicated server.

*A. Network Topology*

For this testbed, a linear topology between the Zodiac FX (OF switch) is presented. The OpenFlow switch S1 will behave similar to the firewall where the OpenFlow rules from the information provided by the CIF will be inserted.

Fig. 1 shows the topology with all the elements connected. The switches are connected to the controller, the FLOW_MOD messages represents the generated rules based on the information provided by the CIF, this information is distributed in multiple tables, which will be responsible for both incoming and outgoing packets.

*B. SDN Controller and Application*

The Opendaylight controller, developed in the Java programming language, is used in this article in its Oxygen-SR2 version. In this work, we made use of the following features:

- L2switch-all: Provides a behaviour of layer 2 switch to the Zodiac FX.
- Dluxapps-applications: It provides a website where it shows the distribution of the nodes.
- Restconf-all: Enables REST interfaces.

We can distinguish two methods for the SDN applications development:

- The development through the consume of the REST API interfaces exposed by the SDN controller, where the selection of programming language is independent of the selected SDN controller, only with support for making REST calls should be considered.
- The development through the implementation of internal service modules MD-SAL, where the libraries and functions of each SDN controller are used, for which the selection of the programming language is conditioned to the chosen SDN Controller.

The choice of one of them depends exclusively on the application to be implemented. For the development of the SDN application in this work, the first method has been selected, where the application is responsible for the creation of the OpenFlow flow entries based on Indicators of Compromise (IOC) provided by the CIF server, and the implementation of these inputs in the switch S1, that fulfills the function of edge switch for our experimental purposes.

*C. Collective Intelligence Framework (CIF)*

The CIF v3, has been installed considering the technical specifications from TABLE I. The CIF periodically receives information reports of known attacks. The most common types of threat intelligence stored in CIF are the IP addresses, domains and URLs, that are observed as related to malicious activities [12], also in the reports, the type of attack is specified in some cases. For this proposal, the information obtained by the CIF is stored in a SQL database. After that, it is filtered because it is possible information is repeated. Once the information is filtered, it is stored in a table to be later used by the application, which through the controller, will be in charge of injecting OpenFlow flow entries that prevent known attacks.

TABLE I. COMPONENTS TECHNICAL SPECIFICATIONS

| Hostname | Operating System | Processor | RAM |
|---|---|---|---|
| CIF Server | Ubuntu 16.04.6 LTS | 2.40 GHz 64-bit Intel Xeon E5-2620 | 188 GB |
| Opendaylight Controller SDN | Ubuntu 16.04.3 LTS | 2.66 GHz Intel Core 2 Quad Q8400 | 8GB |
| Hosts | Ubuntu 16.04.3 LTS | 2.10 GHz 64-bit Intel Core 2 Quad T8100 | 4GB |

## IV. IMPLEMENTATION & RESULTS

Implemented the testbed, the SDN application is executed and three OpenFlow tables are generated, as shown in Fig. 2. Table 0, contains the OpenFlow flow entries that allow communication between hosts within the network. Table 1, contains the OpenFlow flow entries based on the information of CIF that allow to drop traffic whose destination is malicious. Finally, in the Table 2 contains the flow entries based on the information of CIF that allow to drop traffic whose source is malicious.

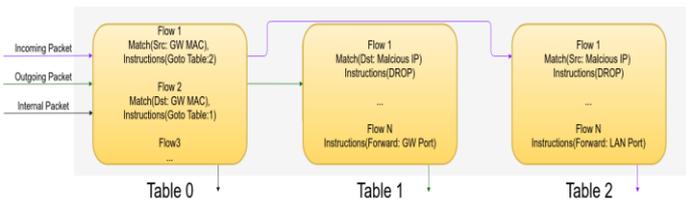

Fig. 2 Flow tables representation.

For communication between hosts, according to Fig. 1, it does not involve switch S1. The packets, whose origin and destination belong to hosts within the same network, arrive at the switch and start at Table 0, which contains flow entries that will indicate which route should packet follow.

In case is required that a host is connect to Internet, the process now involves switch S1. If packets have the gateway as first hop, exists a flow entry in Table 0 of switch S1, which forwards the packets to the Table 1. This table has the flow entries generated by the information provided by the CIF, which have a high priority to ensure that the packets do not make a match with a general rule. The packets makes a sweep of all those flow entries that contain malicious IPs. If there is a match, the packets will be drop; otherwise, the packets will follow its normal course towards the gateway that will generate its subsequent Internet access.

In case the packets come from Internet to the internal network, the process is similar to the previous one but in an inverse way. The packets arrive from the gateway to switch S1. If the packets have as origin the gateway MAC address, there is a flow entry that indicates them go to Table 2.

In Table 2 there are flow entries generated by the information provided by the CIF, which has a high priority to ensure that packets do not make a match with a general flow entry. The packets makes a sweep by all the flow entries and in case of coinciding with a malicious IP, they will be drop. In case they do not coincide with any IP, they will be forward towards the internal network.

In this case, we execute the application that will generate three tables, as shown in Fig 3. We see that in this case, 44 flow entries are generated in Table 1 and Table 2, which are the tables containing information from CIF.

```
Table: 0
 Flows: 7
 Lookups: 394
 Matches: 390
 Bytes: 98687

Table: 1
 Flows: 44
 Lookups: 35
 Matches: 35
 Bytes: 8229

Table: 2
 Flows: 44
 Lookups: 39
 Matches: 39
 Bytes: 14623
```

Fig. 3 Generated Flow tables.

The flows generated in Table 0 are those that allow communication between hosts of the same LAN and at the same time provide the necessary rules to communicate to and from Internet. If the communication goes to or comes from Internet, the packets go to Table 1 or Table 2 respectively.

To verify the operation, we test connectivity from a host to a malicious IP using the ping command, whose rule contain the malicious IP in the Match field and the instruction drop in the Action field. It is observed that initially the flow entry has the parameters Packet Count and Byte Count equal to 0, as shown in Fig 4.

```
Flow 95
 Match:
  ETH Type: IPv4
  Destination IP: ███████
 Attributes:
  Table ID: 1                          Cookie:0x0
  Priority: 65000                      Duration: 556 secs
  Hard Timeout: 0 secs                 Idle Timeout: 0 secs
  Byte Count: 0                        Packet Count: 0
  Last Match: 00:09:16
 Instructions:
  Apply Actions:
   DROP
```

Fig. 4 Initial values flow 95.

Using the ping command, we generate 1000 packets towards the malicious IP (it is omitted). In Fig. 5, it is seen that there is a 100% packet loss, which indicates that no connection could be established.

```
PING ███████ (███████) 56(84) bytes of data.
--- ███████ ping statistics ---
1000 packets transmitted, 0 received, 100% packet loss, time 1001190ms
```

Fig. 5 Failed connectivity test due to the flow entry.

In addition, Fig. 6 shows that the flows statistics have changed, showing now that Packet Count is equal to 1000 and Byte Count is equal to 98000. In that way, we can notice that in case some type of traffic comes from a malicious IP or going to it, this will be blocked by the flow entry on switch S1.

```
Flow 95
Match:
  ETH Type: IPv4
  Destination IP:
Attributes:
  Table ID: 1                          Cookie:0x0
  Priority: 65000                      Duration: 2070 secs
  Hard Timeout: 0 secs                 Idle Timeout: 0 secs
  Byte Count: 98000                    Packet Count: 1000
  Last Match: 00:00:51
Instructions:
  Apply Actions:
    DROP
```

Fig. 6. Flow entry after connectivity test.

## V. CONCLUSIONS

This paper presents a case of use of cyber incidents reports obtained through the CTI management platform (CIF). The data network was programmed through the successful development of an SDN application using Python language, which translates the incident reports into OpenFlow flow entries to be inserted in the table corresponding of OpenFlow switch when the pipeline process is enabled. The testbed was made up of OpenFlow switches for domestic use, which are not designed to work in a production environment, due to the limited number of ports available, however this did not mean an impediment to be able to carry out the validation process successfully.

As future work, we propose to add the classification mechanisms of incident records obtained from internal elements of a network, in order to obtain the threat intelligence lifecycle.

## VI. ACKNOWLEDGMENT

We would especially like to thank Rodolfo Ruiz B. for his valuable contribution to INICTEL-UNI by donating the HITACHI UCP-2000 infrastructure, used for the implementation of cyber threat intelligence. This work was development in the network and cybersecurity research laboratory of INICTEL-UNI.